\documentstyle[12pt,epsfig,aaspp4]{article}
\lefthead{Aparicio et al.}
\righthead{The Stellar Populations of Antlia}
\def\surfb{{\rm\,mag/(\arcsec)^2}}
\def\kpc{{\rm\,kpc}}
\def\msun{{\rm\,M_\odot}}
\def\kms{{\rm\,km/s}}
\begin{document}

\title{The Nature of the New Local Group Dwarf Galaxy Antlia\altaffilmark{1}}

\author{A. Aparicio\altaffilmark{2}}

\affil{Instituto de Astrof\'\i sica de Canarias, E-38200  La Laguna, Tenerife, Canary Islands, Spain. e-mail: aaj@iac.es}

\author{J. J. Dalcanton\altaffilmark{3}}

\affil{Observatories of the Carnegie Institution of Washington, 813 Santa Barbara St., Pasadena, CA 91101, USA. e-mail: jd@ociw.edu}

\author{C. Gallart}

\affil{Observatories of the Carnegie Institution of Washington, 813 Santa Barbara St., Pasadena, CA 91101, USA. e-mail: carme@ociw.edu}

\and

\author{D. Mart\'\i nez-Delgado}

\affil{Instituto de Astrof\'\i sica de Canarias, E-38200  La Laguna, Tenerife, Canary Islands, Spain. e-mail: ddelgado@iac.es}

\altaffiltext{1} {Based on observations made with the 2.5 m Du Pont telescope of the Carnegie Institution of Washington at Las Campanas Observatory in Chile.}

\altaffiltext{2}{Presently at the Observatories of the Carnegie Institution of Washington, 813 Santa Barbara St., Pasadena, CA 91101, USA} 

\altaffiltext{3}{Hubble Fellow}

\begin{abstract}

The recently discovered Antlia dwarf galaxy is analyzed using $VI$
photometry. The galaxy is resolved into a large number of stars and
although the galaxy is intrinsically faint and low surface brightness,
its stellar populations reveal characteristics more typical of faint
star-forming dIrs rather than dEs. Significant star formation is
currently going on in the central part of Antlia although little or no
star formation is taking place in the outer regions. This indicates a
two-component (core-halo) morphology which appears to be common, not only in large spirals (disk-halo), but in dwarf galaxies as well. The SFR averaged over the lifetime of the galaxy is estimated to be $\bar\psi/A\sim 2-4\times
10^{-10}$ M$_{\sun}$yr$^{-1}$pc$^{-2}$ while the more recent star
formation, averaged over the last 1 Gyr is much higher ($\bar\psi_{\rm
1Gyr}/A\sim 3-9\times 10^{-10}$ M$_{\sun}$yr$^{-1}$pc$^{-2}$ for the
central region). The total mass locked into stars and stellar remnants
is estimated to be $M_\star\sim 2-4\times 10^6$ M$_{\sun}$. Its
distance, estimated from the TRGB is $1.32\pm 0.06$ Mpc, which places
Antlia just beyond the Local Group, and makes it a close companion of
the dwarf galaxy NGC 3109 ($\Delta r\gtrsim30\kpc$), although it is not clear whether they are gravitationally bound.

\end{abstract}

\section{Introduction} \label{intro}

Dwarf galaxies are probably the most common type of galaxies in the
Universe. At the low end of the mass and luminosity distributions,
they span a very wide range of properties and morphologies. Dwarf
galaxies can be divided into two large groups: dwarf irregulars (dIrs)
and dwarf ellipticals (dEs) (we include in this group the so-called
dwarf spheroidal galaxies). In the simplest scheme, the key
differences between the two groups are that dIrs have large amounts of
gas and high current star formation activity producing many HII
regions and blue stars, while dEs lack gas and contain only old
($>10$ Gyr) stars\footnote {In this paper, we mean by old
stars those older than 10 Gyr, by intermediate-age stars those with
ages between 1 and 10 Gyr and by young stars those younger than 1
Gyr.}. Morphologically, dIrs tend to be patchy small blue galaxies like
NGC 6822 or IC 1613, while dEs show a smooth luminosity distribution
(e.g.\ NGC 147) and in many cases are hardly distinguishable from the sky
background, like Sextans or the small Andromeda companions.

However, this picture has been shown to be too simple and the
separation between both groups is quite diffuse, if any. On the one
hand, there are objects with large amounts of gas but showing an
apparently low star formation rate (SFR), such as Pegasus (Aparicio,
Gallart, \& Bertelli \markcite{peg} 1997a) or LGS 3 (Aparicio,
Gallart, \& Bertelli \markcite{lgs} 1997b). These systems have few or
no HII regions and lack the patchy morphology of more active
dIrs. But, since these galaxies are equally gas rich, the
different appearances among dIrs may be the result of episodic
star formation, in such a way that some of them are in an
active phase of star formation while others are currently quiescent.
On the other hand, growing evidence exists that objects traditionally
considered to be dEs have a substantial intermediate age
stellar population. This was first shown by the discovery of Carbon
stars in most of the dE satellites of the Milky Way (see
Azzopardi\markcite{azzopardi} 1994 and references therein) and then
beautifully shown in the deep color-magnitude (CM) diagrams obtained
for some of these galaxies, like Carina (Smeker-Hane, Stetson, \& Hesser\markcite{smeker} 1994), Leo I (Lee et al.\markcite{leoi} 1993a),
Fornax (Stetson\markcite{fornax} 1997) and Phoenix
(Mart\'\i nez-Delgado, Aparicio, \& Gallart\markcite{phoenix} 1998). There
is also some evidence of an intermediate age population in the larger
dE Andromeda companions, particularly in the cases of NGC~185
(Mart\'\i nez-Delgado, Aparicio, \& Gallart\markcite{n185} 1997) and
NGC~205 (Davidge\markcite{davidge92} 1992). Therefore it is not clear
where the true dividing line between dIrs and dEs (if any) should be
placed; even the gas content, which should tell us about a galaxy's
potential to form stars is not a clear indicator, since galaxies which are
apparently devoid of gas, such as Fornax (Stetson 1997), somehow manage to keep
forming stars. All this calls to a revision of our current
understanding of the nature of dwarf galaxies.

In order to further explore the differences among the dwarf galaxy
population, we have undertaken a study of the newly rediscovered
Antlia dwarf galay.  The galaxy had been noted by Crowin et al.\markcite{corwin} (1985); Feitzinger \& Galinsky\markcite{feitz} (1985) and Arp \&
Madore\markcite{arp} (1987), but was overlooked for a decade. Whiting, Irwin \& Haw\markcite{whitin} (1997) rediscovered the galaxy from inspection of the UK Schmidt Telescope plates covering the entire southern sky. Deep CCD images on the 1.5m telescope at Cerro Tololo Observatory performed by these authors revealed it completely resolved into individual stars, indicating that it was probably a new Local Group member. From wide band and H$_\alpha$
images, these authors observed no hot blue stars nor any obvious trace
of star forming regions. They concluded that the galaxy is a typical
dE similar to the Milky Way satellites and to the Tucana dwarf. They
estimate the distance to be about 1.15 Mpc and its diameter 1--2 Kpc.

We have performed deep $VI$ photometry of the galaxy resolving a large
number of stars. In contrast to Whiting et al. (1997), our results show
that star formation is currently taking place (or stopped just a few
Myr ago) in the central part of Antlia, making it similar to the
smallest, faintest dIr galaxies like Pegasus, LGS 3 or Phoenix. The
paper is organized as follows. Section \ref{obs} briefly presents the
observations. In \S \ref{cmmain} an overview of the CM diagram is
given and used to estimate the distance to Antlia and its
metallicity. In \S \ref{light} the integrated light distribution is
discussed. The star formation history (SFH) is sketched in \S
\ref{sfh}, through comparison to other well studied faint
dIrs. Finally, the results are summarized in \S \ref{conclusiones}.

\section{Observations, data reduction and photometry} \label{obs}

Observations of the Antlia dwarf were made at Las Campanas
Observatories on the nights of April 10-12 and April 14, 1997 under
photometric conditions, using the Dupont 2.5m telescope with a
2048$\times$2048 thinned Tektronics chip with 0.259\arcsec pixels, a
gain of 2.2 e$^-$/adu, and a read noise of 6.6 e$^-$.  The journal of
observations is given in Table 1.

All images were bias subtracted using the overscan region, and
corrected for residual 2-dimensional structure using a bias frame.
Dome flats were used to remove pixel-to-pixel and large scale variations in the chip response. A fringe frame was created from a median all of the images taken on the night of April 14th of high latitude nearly-blank fields and was subtracted from all $I$ band images, after matching the sky levels.  The sky in the resulting frames is flat to within 0.7\%.

Photometry of the stars in Antlia has been performed using the new
DAOPHOT/ALLSTAR/ALLFRAME software, made available to us by Dr. Stetson
(see Stetson\markcite{allframe} 1994). A master frame has been
produced using the best seeing images, both $V$ and $I$, obtained
throughout the run. The total integration times are 2400 sec in $V$
and 2400 sec in $I$ and the FWHM of the stellar images is 0.8\arcsec
~for the combined image. Three FIND/ALLSTAR/SUBSTAR passes have been
used to identify stars in this master frame. The resulting list,
consisting on 6831 stars, has been given to ALLFRAME to perform the
photometry in the 16 individual frames using quadratically varying
PSFs computed for each individual frame. The $V$ and $I$ magnitudes
for the stars have been obtained as the robust average of the
magnitudes obtained in the individual frames. Only stars that have
been measured in at least three images have been considered. The
resulting $V$ and $I$ photometry lists have been paired using
DAOMASTER. A final list of 5566 stars is the result of this
process. This catalogue of stars has been filtered using the error
parameters given by ALLFRAME, retaining only stars with acceptable CHI
and SHARP parameters and $\sigma_{\rm V}\leq 0.2$ and $\sigma_{\rm
I}\leq 0.2$.

Aperture corrections have been obtained using a large number of well
measured stars in one $V$ and one $I$ image of Antlia. They are
accurate to $\pm 0.005$ mag in $V$ and $\pm 0.008$ mag in
$I$. Extinction for each night and the final transformation to the
Johnson-Cousins standard system has been computed using several
standards in the list of Landolt\markcite{landolt} (1992) observed
repeatedly throughout each of the nights. Color terms have been found
to be comparable to the dispersion of the data and have not been used. The
standard errors of the extinctions are always smaller than 0.01, while
the standard errors in the zero points are $\pm 0.005$ magnitudes in
$V$ and $\pm 0.006$ magnitudes in $I$. The total errors in the
zeropoint of the photometry are therefore about 0.01 in both $V$ and
$I$.

Figure 1 shows the $I$ image of Antlia. For the analysis of
the stellar content (see \S\ref{cmmain} and \S\ref{sfh}), the observed
field has been divided into several parts. Region A corresponds to the
central part of Antlia; region B covers the optical external part of
the galaxy and regions C map the foreground field stars. A finding
chart of the resolved stars with well measured photometry is shown in
Fig. 2 (see \S\ref{foreground}). The central part is
displayed in Fig. 3, which shows the beautiful resolution
into stars of the central part of Antlia. The circle indicates a
possible HII region in the center of this region, which corresponds to
a large concentration of blue stars. That this central object is
actually an HII region needs to be confirmed with deep $H_\alpha$
observations.

\section{The Color-Magnitude Diagram, the Metallicity and the Distance} \label{cmmain}

\subsection{The Color-Magnitude Diagram: Subtraction of Foreground Population and Reddening} \label{foreground}

The CM diagrams of regions A, B and C (see Fig. 1) are shown
in Fig. 4. Panels [a] and [b] show the CM diagrams of the inner
(A) and outer (B) galaxy. Panels [c] and [d] are CM diagrams of the
external (C) regions. They show only a randomly chosen fraction of the
stars in these regions such that sampled areas are the same as those
of regions A and B, respectively; no attempt has been made to correct
for the difference in crowding between the galaxy and field
regions. These diagrams have been used to remove foreground stars from
diagrams [a] and [b]. To do so, for each star in the foreground frames
([c] \& [d]), we select all the stars of the corresponding galaxy
frame which lie within a range of magnitude and color around the
selected foreground star. This range is the ALLFRAME
error corresponding to the magnitude of the foreground star. In a
second step, one of the selected stars is randomly chosen and
removed. If no stars happen to be inside the error interval, then the
nearest one within 0.5 magnitudes is removed. The resulting clean
diagrams are plotted in panels [e] and [f]. Field subtraction in the
central region (A) is quite independent on the particular subsample of
field stars used, due to the low percentage of foreground stars.  This
is not the case for the low surface brightness outer region (B), and
particularly affects the blue stars and those above the TRGB.  Thus,
it cannot be ruled out that the few stars that remain in these regions
in the outer galaxy result from bad field subtraction. Foreground
stars as well as those removed from regions A and B are plotted with
small dots in Fig. 2. The open circles mark the stars that
have been left in the CM diagrams as galaxy members after the field
subtraction, and the filled triangles mark blue stars from panels [e]
and [f] in Fig. 4.

The Antlia $[(V-I),I]$ CM diagram shows characteristics typical of the
CM diagrams of other dIr galaxies, with evidence for both an old and a
young population. The old and intermediate age population clump
together in what we are calling the {\it red-tangle} (see Aparicio, \&
Gallart\markcite{lg} 1994), which corresponds to the RGB and AGB
locii.  The red tangle is the most prominent feature of the Antlia CM
diagram, running from $[(V-I),I]\simeq [1.0,24.0]$ to
$[1.5,21.5]$. There is also a considerable population of blue stars
($(V-I)\lesssim 0.7$) in the CM diagram of the central region (panel
[e] of Fig. 4). It is remarkable that these blue stars are
entirely absent from the CM diagram of the outer part of the galaxy
(panel [f]). The presence of these stars in the central region, as
well as the candidate HII region (Fig. 3), support the idea
that Antlia is a dIr galaxy rather than a dE (see \S\ref{sfh}). Most
of the bright stars ($I\lesssim 21$) are likely foreground stars.

The diagrams shown in Fig. 4 have not been corrected for interstellar extinction. However, the extinction must be considered in the forthcoming calculation of metallicity and distance. The galactic coordinates of Antlia are $l_{\rm II}=263^{\rm o}.09$, $b_{\rm II}=22^{\rm o}.32$, which lies in a
region of moderate gradient and between the isolines corresponding to
$E(B-V)=0.03$ and $E(B-V)=0.06$ in the extinction maps of Burstein \& Heiles\markcite{burstein} (1982). Neglecting internal reddening, we
assume $E(B-V)=0.045\pm 0.015$. Using the extinction law of Cardelli,
Clayton, \& Mathis\markcite{cardelli} (1989) with $R_{\rm V}=3.3$, the
extinctions in $V$ and $I$ are $A_{\rm V}=0.15\pm 0.05$ and $A_{\rm
I}=0.07\pm 0.03$.

\subsection{Metallicity} \label{metalicidad}

The average stellar metallicity can be obtained from $(V-I)_{\rm
-3.5}$, which is the color index of the RGB half a magnitude below the
TRGB (see Da Costa \& Armandroff \markcite{dacosta} 1990; Lee, Freedman, \& Madore \markcite{trgb} 1993b). We have estimated $(V-I)_{\rm -3.5}$ from
the median of the stars with colors in the interval $1.2<(V-I)\leq
1.8$ and with magnitudes in the interval $22.04<I\leq 22.24$ (the
magnitude of the TRGB is $I_{\rm TRGB}=21.64$; see
\S\ref{distancia}). We find $(V-I)_{-3.5}=1.45$, with a color
dispersion of $\pm 0.07$. Similar values are obtained using stars in
the outer region B: 1.46 and 0.08, respectively. The photometric error
yielded by ALLFRAME at this magnitude is about $\pm 0.025$. As shown
by Aparicio \& Gallart \markcite{pegfot} (1995) and Gallart, Aparicio,
\& V\'\i lchez\markcite{n6822a} (1996a), the actual external errors
affecting the photometry cannot be in general estimated from this and
are quite dependent on other observational effects (crowding, blending
and others). However, the fact that we obtain quite similar values for
the color dispersion in central and outer regions indicates that these
effects are small and that probably, most of the observational
dispersion is due to signal-to-noise at these magnitudes. In these
conditions, ALLFRAME errors may be a reasonable estimate of the actual
total errors. Furthermore, considering that at the magnitude we are
dealing with, the photometric uncertainty is small compared with the
color dispersion we measure, it is reasonable to assume that the
latter is a good representation of the intrinsic dispersion. We will
use it below to estimate the dispersion in metallicity. 

The resulting dereddened value for the color index of the RGB is
$(V-I)_{\rm -3.5,0}=1.37$. Using the calibration by Lee et
al. (1993b), we obtain a metallicity $[Fe/H]=-1.6\pm 0.1$,
corresponding to $Z=0.0012\pm 0.0003$ and in good agreement with the estimate by Whiting et al. (1997). The error has been obtained
from propagation of the reddening error. Under the assumption that the
color dispersion of the red-tangle is only produced by metallicity
dispersion, the latter is $-1.9<[Fe/H]<-1.4$.  For comparison, Fornax
has a metallicity of $[Fe/H]=-1.4$ and Phoenix has a metallicity of
$[Fe/H]=-1.9$.

\subsection{Distance} \label{distancia}

The absolute $I$ magnitude of the tip of the RGB (TRGB) has proven to
be a remarkably good estimator of distance (Lee et al. 1993b).  
The absolute magnitude of the TRGB is slightly
dependent on the metallicity, which we have estimated above in
\S\ref{metalicidad}. To obtain the magnitude of the TRGB we have used
the luminosity function (LF) of the stars in the color interval
$1.0<(V-I)\leq 2.0$ drawn from the central galaxy (region A), after
correction for foreground contamination (diagram [e] in
Fig. 4). We have computed the LF by counting the stars lying
inside an interval of $\pm 0.2$ magnitudes about a central value of
$I$. The central value has been varied in steps of 0.02 magnitudes to
obtain the LF while reducing the dependence of the results upon the
particular choice of bin center. Finally, an edge detecting Sobel
filter $[-1,0,+1]$ has been aplied to the LF. This produces a sharp
peak at the TRGB corresponding to $I_{\rm TRGB}=21.64\pm 0.04$. The
error has been obtained from the width of the peak at 1/3 of its
heigh. The same process applied to the foreground stars alone yields a
noisy distribution with a dispersion which is 3.5 times smaller than
the height of the peak detected for region A. The resulting dereddened
value is $I_{\rm TRGB,0}=21.57\pm 0.05$. The error has been calculated
from the quadratic addition of the Sobel filter and extinction errors.

The color index of the TRGB is necessary to calculate the bolometric correction to be used in the $I_{\rm TRGB}$-distance calibration. It has been estimated from the median color index of the stars with $1.2<(V-I)\leq 1.8$ and $21.64<I\leq 21.84$. We find $(V-I)_{\rm TRGB}=1.50$ which when corrected for external extinction yields $(V-I)_{\rm TRGB,0}=1.42$.

Using the calibration by Lee et al. (1993b), we obtain a distance
modulus $(m-M)_0=25.6\pm 0.1$, corresponding to $1.32\pm 0.06$
Mpc. The intrinsic error of the method is about $\pm 0.1$ (see Lee
et al. 1993a). This is the adopted error since it is larger than the
photometric and extiction ones. Although marginally compatible, our distance is slightly larger than the result by Whiting et al. (1997). The reason for the disagreement might be an upward shift of the TRGB in Whiting et al. CM diagram due to crowding effects at the level of their limiting magnitude.

Antlia lies $\sim 1.2^{\rm o}$ South from NGC 3109 and both galaxies
lie at the same distance, inside the error intervals (see Capaccioli,
Piotto \& Bresolin\markcite{capa} 1992; Lee\markcite{lee} 1993;
Davidge\markcite{davidge} 1993). Assuming the same distance for both
galaxies, their minimum physical separation is $r=29$ Kpc.  However, if the distance of NGC 3109 is assumed to be $1.26\pm0.1$ Mpc (Capaccioli et al. 1992; Lee 1993), and the errors are considered, the true separation
could be as large as 180 Kpc. In spite of their close proximity, the difference in the velocities makes not clear whether Antlia is  gravitationally bound to NGC 3109. Assuming a mass of $8.6\times 10^9\msun$ for NGC 3109 (Jobin \& Carignan 1990),  and circular orbits, Antlia would be unbound if the relative velocity exceeded only $51\kms\times (r/29{\rm Kpc})^{-1/2}$. The radial velocity of NGC 3109 is 406 Kms$^{-1}$ (Hartmann\markcite{hartmann1} 1994 and\markcite{hartmann2} 1997), while that of Antlia is 361 Kms$^{-1}$ (Fouqu\'e et al.\markcite{fouque} 1990). This means that the galaxies would be unbound if the realtive distance is $r\gtrsim 37$ Kpc. Considering that still smaller values will be in general expected for non-circular orbits, it is more likely that either they are both bound to the larger mass of the Local Group (although van den Bergh \markcite{vanden}1994 places NGC 3109 just beyond the gravitational limit of the Local Group), or they are members of a common filament feeding into the Local Group region.

\section{Integrated Light} \label{light}

To determine the total magnitude, color, and surface brightness
distribution of Antlia, we have fit its surface brightness
distribution with ellipses of ellipticity $e=1-b/a=0.4$ and semi-major
axis at position angle $135^{\rm o}$ (measured from North to East), in
both the $V$ and the $I$ bands, after masking out stars which were too
bright to be members of the galaxy (e.g.\ $I<21$).  The sky was
subtracted using the median surface brightness distribution in the
ellipses beyond a radius of $2\arcmin$.  

The resulting $V$-band surface brightness distribution is shown in
Figure 5 as a function of the ellipse semi-major axis $a$.
In both the $V$ and $I$ bands, the surface brightness distribution is
fairly flat to a radius of $\sim\!40{\arcsec}$, which corresponds
roughly to the extent of the central galaxy defined by region (A), and
then falls off more steeply at larger radii.  This is comparable to
the profiles of the faint Andromeda companions And I, And II, \& And
III, reported by Caldwell et al.\ (1992).  The $V$ band surface
brightness can be well fit by a double exponential profile with
central surface brightness $\mu_0(V)=24.3\surfb$ and exponential scale
length $\alpha=64{\arcsec}$ in the inner regions ($a\!<\!40{\arcsec}$),
and $\mu_0(V)=22.4\surfb$ and $\alpha=16.5{\arcsec}$ at larger radii.
This gives an angular diameter at the $\mu_{\rm V}=26\surfb$ isophote of
$\theta_{26}=1.8{\arcmin}$; this corresponds to an isophote of
$\mu_{\rm I}=25.1\surfb$.  Given the distance of the galaxy, the break in
the surface brightness profile occurs at $a=0.26$ Kpc, and the inner
and outer scale lengths are 0.36 Kpc and 0.14 Kpc, respectively.

Integrating the surface brightness profiles within the $V$ and $I$
bands out to the $\mu_{\rm V}=26\surfb$ isophote, we find
$m_{\rm V}\!=\!15.67\pm0.12$, and $m_{\rm I}\!=\!14.91\pm0.12$, uncorrected for
internal or galactic extinction.  The quoted errors are an upper limit
computed assuming that every point in Fig. 5 were
systematically off by 1 $\sigma$. Following de Vaucouleurs et al. (1991), the internal extinction of Antlia can be estimated to be $A_{\rm B,i}=0.30$. Using the extinction law by Cardelli et al. (1989) with $R_{\rm V}=3.3$ this yields $A_{\rm V,i}=0.23$ and $A_{\rm I,i}=0.11$. After correction for internal and external extinction and considering the error in the distance modulus, the total absolute magnitude of Antlia is $M_{\rm V,0}=-10.3\pm 0.2$ and $M_{\rm I,0}=-10.9\pm 0.2$.  The luminosity of Antlia is therefore
comparable to the faintest dIrs, making Antlia an extremely faint
galaxy in spite of its recent star formation.  The mean color of
Antlia is quite blue: $(V-I)_0\!=\!0.6\pm0.2$, 
and is comparable to the colors of galaxy types
Scd \& Im ($(V-I)\!=\!0.82$) for the Coleman et al.\markcite{coleman} (1980) composite
spectral energy distributions, and Im ($V-I\!=\!0.64$) for the SEDs reported by Fukugita et al.\markcite{fukugita}
(1995).

\section{Clues to the Star Formation History} \label{sfh}

The color magnitude diagrams presented in \S\ref{cmmain} clearly show
that Antlia has had an extended star formation
history, as evidenced by its old and intermediate age red-tangle
and its population of young, blue stars.  We can sketch
the SFH of Antlia in more detail by comparing Antlia with two faint,
low surface brightness dIrs whose CM diagrams are similar to Antlia's
and whose star formation histories have been well studied:
Pegasus (Aparicio et al. 1997a) and LGS 3 (Aparicio et al. 1997b). The
SFHs of these galaxies have been determined from their resolved stellar
content (former references), hence from information similar to the one
available for Antlia. It turns out, as we will see below, that Antlia
has characteristics intermediate between these two galaxies.

Figure 6 shows the CM diagrams of the central and outer
regions of Antlia compared with those of Pegasus and LGS 3. An
isochrone from the library of Padua (see Bertelli et
al.\markcite{bertelli} 1994) of 0.1 Gyr and $Z=0.001$ for Antlia and
LGS 3 and the same age and $Z=0.004$ for Pegasus, is overplotted in
the four diagrams. The isochrone qualitatively shows that the blue
stars present in the CM diagrams of the central region of Antlia are
likely young main-sequence (MS) and/or blue-loops (BL) stars a few
hundred Myr old, indicating that recent star formation activity has
taken place in the central part of the galaxy. The few stars observed
above the TRGB (within $\sim 1$ magnitude) in Antlia's central CM
diagram are likely intermediate-age AGB stars, similar to what are
observed in Pegasus and, to a lesser extent, in LGS 3. These stars
indicate star formation activity at intermediate ages. A few of these
blue stars and likely intermediate age stars are also seen in the outer
regions of Antlia, but, as stated above, they may also result from
incomplete field subtraction in the more highly contaminated outer
regions. The fact that Antlia lacks an extended AGB red tail structure
like Pegasus (seen redward of $(V-I)\!=\!2$, at $M_{\rm I}\!=\!-3.5$), is
consistent with its low metallicity (Aparicio \& Gallart 1994).

If star formation has been very recent in the central region, HII
regions would be expected. However, conspicuous HII regions are not
necessarily required if the star formation is going on at a slow
rate. This is also the case of both Pegasus and LGS 3. Interestingly,
a feature is seen in our images of Antlia that could be a small HII
region. It is marked with a circle in Fig. 3 and its
position corresponds to the clump of blue stars at $(x,y)\simeq
(700,950)$ shown in Fig. 2. If the feature proves to be an
HII region, it would be definitive evidence of significant star
formation activity in the last $\sim 10$ Myr.

The SFHs of Pegasus (Aparicio et al. 1997a) and LGS 3 (Aparicio et
al. 1997b) have been obtained from comparison of their CM diagrams
with a large set of model CM diagrams computed for different SFRs and
chemical enrichment laws. We will compare the CM diagram of Pegasus
and LGS~3 with that of Antlia to provide an insight to its
SFH. Gallart et al. (1996b) showed that the distributions of stars
older and younger than $\sim 1$ Gyr are decoupled in a CM diagram like
that of Antlia: almost no stars older than 1 Gyr are among the blue
population and almost no stars younger than that age populate the
red-tangle region. On the other hand, the amount of blue and
red-tangle stars should be related to the SFR $\psi(t)$, integrated
over the appropriate time intervals. Therefore the number of blue and
red stars in Antlia, compared to that in Pegasus and LGS 3, will
constrain Antlia's SFH. The results are summarized in Table 2 where
the number of red and blue stars, the ratio between them and the SFR
averaged for the whole life of the galaxies ($\bar\psi$) and for the
last 1 Gyr ($\bar\psi_{\rm 1Gyr}$) are given.  The SFR values for
Antlia are given in brackets, indicating that they are estimates from
comparison with the other two galaxies, derived by scaling the SFR in
LGS 3 and Pegasus to the relative number of blue and red stars. Data
for the inner (A) and outer (B) regions of Antlia are considered
separately. Only the stars between the TRGB and 1.5 magnitudes in $I$
below it have been considered. To separate the stars into blue and red
categories, the median $(V-I)$ of the RGB at $I=I_{\rm TRGB}+1.5$ has
been used as reference. The stars are counted as blue when they are at
least 0.5 magnitudes bluer than that $(V-I)$; otherwise, they are
counted as red.

The numbers of red and blue stars listed in Table 2 suggest that both
the recent and the past SFR in the central part of Antlia, have been
intermediate between those of LGS 3 and Pegasus.  In the center of
Antlia, the galaxy has been forming stars for the last 1 Gyr at roughly twice its normal
rate, whereas in Pegasus, the recent star formation is comparable to
the average lifetime SFR, and in LGS3, the recent SFR is a factor of
three smaller than its lifetime average.  In the outer regions of
Antlia, it seems that the past SFR was much higher than it has been in
the most recent 1 Gyr.  However, the small upper limit given for the
current SFR in the outer region is highly uncertain because of the
uncertainties of the foreground stars subtraction in the more heavily
contaminated outer galaxy region.

Table 3 summarizes the SFR normalized by the galaxies' area and some
integrated physical parameters for Antlia, Pegasus, and
LGS 3. The areas of Antlia regions A and B are 3.06 $(\arcmin)^2$ and
11.29 $(\arcmin)^2$ respectively, which correspond to $4.51\times
10^5$ pc$^2$ and $1.66\times 10^6$ pc$^2$, respectively. The values
for Pegasus (see Aparicio et al. 1997a and references therein) and LGS
3 (see Aparicio et al. 1997b and referencies therein) are also given
for comparison. The distance of Antlia from the barycenter of the Local
Group has been calculated assuming the latter is 0.45 Mpc from
the Milky Way towards Andromeda (see Aparicio et al. 1997b and
references therein). To calculate $M_{\rm V,0}$, the integrated $V_{\rm T}$ magnitude of Pegasus and LGS 3 given respectively by de Vaucouleurs et al.\markcite{dvau} (1991) and by  Karachentseva et al.\markcite{kar} (1997) have been used. The absolute magnitude of the Sun has been assumed to be $M_{\rm V\sun}=4.87$ (Durrant\markcite{durrant} 1981). The total mass in stars and stellar remnants ($M_\star$) of Antlia has been calculated from $\bar\psi$ and
$\bar\psi_{\rm 1Gyr}$ for the central and outer regions. The value of the gas mass ($M_{\rm gas}$) has been calculated using the data by Fouqu\'e et al. (1990); the relationship given by Hoffman et al.\markcite{hoffman} (1996) for the HI mass, and assuming that $M_{\rm gas}=\slantfrac{4}{3} M_{\rm HI}$. The total mass ($M_{\rm tot}$) has been estimated using the relation $M_{\rm tot}=9.3\times 10^5 \bar r_{\rm H} \sigma^2$ given by Lo, Sargent, \& Young\markcite{lo} (1993), which assumes virial equilibrium. $\bar r_{\rm H}$ has been obtained as $(a\times b)^{1/2}$, where $a=0.7$ Kpc is the semimajor axis of the galaxy at the $\mu_{\rm V}=26\surfb$ isophote; $b=0.4$ is the semiminor axis, calculated using the ellipticity $e=0.4$ (see sec. \ref{light}) and $\sigma=9$ Km s$^{-1}$ is the velocity dispersion estimated from the value of $W_{50}$ given by Fouqu\'e et al. (1990), assuming a gaussian profile of the 21 cm line.

The significant recent star formation observed in the central region
of Antlia contrasts with the absence of a significant young population in the
outer region and may suggest a two components (core-halo or disk-halo) composition of the galaxy. This seems to be common not only in large spirals but also in dwarfs (see for example the case of WLM; Minniti \& Zijlstra\markcite{minniti} 1996) indicating that the gas currently participating in the star formation is concentrated in the central part of the galaxy and that
the physical mechanisms of galaxy formation are shared by systems in a
large range of initial masses. Nevertheless at the distances of the
dIr galaxies, it is difficult to ascertain whether what we are
observing as a "halo" consists of a purely "old" (age $>10$ Gyr)
population, or that at the low densities in the outer regions, star
formation procedes much more slowly than in the dense center, such
that the quantity of young stars at any time is much smaller than in
the central region, and hence virtually unobservable. Note that the
structure of the red-tangle in both the central and the outer region
is very similar, suggesting similar SFH from 1 Gyr to 15 Gyr for both
parts of the galaxy. Although appealing, we argue that this
hypothetical, truly dual, core-halo structure in dIr galaxies should
be further investigated. The analysis of the stellar population in a large enough area containing both kind of (core-halo) population, through comparison with model CM diagrams (see Aparicio et al. 1997a and 1997b and Gallart et al. 1997b) may cast some light to this issue.

\section{Conclusions} \label{conclusiones}

We have presented photometry of resolved stars in the recently
discovered Antlia dwarf galaxy. Although it is an intrinsically faint,
low surface brightness galaxy, Antlia has characteristics of dIr
galaxies rather than dEs, at least from the point of view of the
relatively large amount of recent star formation detected in its
central region. Its HI mass is also compatible with a dIr galaxy. 
Future measures of its HI and HII contents will help
towards understanding its properties. 

The stellar component of the galaxy has an average metallicity
$[Fe/H]=-1.6\pm 0.1$, corresponding to $Z=0.0012\pm 0.0003$, simlar to
the less enriched dIrs and dEs.  This places Antlia nicely on the
relationships $[Fe/H]-M_{\rm V}$ and $[Fe/H]-\mu_{\rm V}$ shown by
Caldwell et al. (1992). Antlia is at a distance of $1.32\pm 0.06$ Mpc,
at what is considered to be the outskirts of the Local Group.  It is
also a close companion ($29\kpc\!\le\! r\!\le\!180\kpc$) of NGC 3109, although  their relative velocity (45 Kms$^{-1}$) makes it unlikely that they form a gravitationally bound pair. They would be bound only if their relative distance is smaller than 37 Kp and the orbit is circular.

Antlia has a light distribution which is well described by a double
exponential, with a relatively flat distribution within $r=40^\prime$
which steepens at larger radii.  Its central surface brightness is
$\mu_0(V)=24.3\surfb$, and its isophotal diameter along the semi-major
axis is $\theta_{26}=1.8{\arcmin}$ or $D_{26}=0.7\kpc$, making it
an intrinsically small low surface brightness galaxy.  Antlia's integrated
absolute magnitude within $\mu_{\rm V}=26\surfb$ is $M_{\rm V,0}=-10.3\pm 0.2$, making
it among the lowest luminosity galaxies known.  It is also quite blue
($(V-I)\!=\!0.6\pm0.2$), and is comparable in color to very late
type Scd/Im galaxies.

The Antlia SFH has been discussed through comparison with the CM
diagrams of Pegasus and LGS 3. Star formation is currently going on
(or has stoped within the last few Myr) in the central $\sim 0.25$ Kpc
of Antlia, at a rate at least twice as large as the average for the
entire galaxy lifetime (see Table 2). This rate is similar to and probably
intermediate between that of Pegasus and LGS 3. Little or no star
formation is currently taking place in the outer region of the
galaxy. This behaviour seems to be quite common in dwarf galaxies and
might be indicative of a core-halo structure in which the gas
participating in the star formation would be concentrated towards the
center of the galaxy. Nevertheless, we argue that the question of
whether the outer part of the galaxy showing little or no current star
formation is or not a {\it true} halo, composed solely by an old
($>10$ Gyr old) population should be further investigated.  Given the
integrated star formation history, we estimate that the total mass
into stars and stellar remnants is $M_\star\sim 2-4\times 10^6$
M$_{\sun}$.

\acknowledgements

We are grateful to Dr. A. Whiting and Dr. M. Irwin for giving us information about the galaxy and for making available to us a preprint of their Antlia paper, and to Dr. W. Freedman for advising us of the previous
discovery of the Antlia galaxy by Arp \& Madore.  We also thank
Dr. D. Hartmann for providing us with HI maps of the Antlia/NGC3109
region.  AA thanks the Observatories of the Carnegie Institution for
its hospitality. AA is financially supported by the Instituto de
Astrof\'\i sica de Canarias (grant P3/94) and by the Direcci\'on
General de Ense\~nanza Superior of the Kingdom of Spain (grant
PB94-0433). Support for JJD was provided by NASA through Hubble
Fellowship grant \#2-6649 awarded by the Space Telescope Science
Institute, which is operated by the Association of Universities for
Research in Astronomy, Inc., for NASA under contract NAS 5-26555. DMD
is financially supported by the Instituto de Astrof\'\i sica de
Canarias (grant P3/94).

\newpage

\begin{deluxetable}{ccccc}
\tablenum{1}
\tablewidth{0pt}
\tablecaption{Journal of Observations\label{journal}}
\tablehead{
\colhead{Date}      &
\colhead{Time(UT)}      &
\colhead{Filter}  &
\colhead{Exp. time (s)}          & \colhead{FWHM ($^{\prime\prime}$)}}
\startdata
97.04.12	& 00:03	& V	& 600	& 1.15	\nl
97.04.12	& 00:16	& V	& 600	& 1.12	\nl
97.04.12	& 00:28	& V	& 600	& 1.14	\nl
97.04.12	& 23:45	& I	& 600	& 0.75	\nl
97.04.13	& 00:04	& I	& 300	& 0.67	\nl
97.04.13	& 00:11	& I	& 300	& 0.75	\nl
97.04.13	& 00:18	& I	& 300	& 0.72	\nl
97.04.13	& 00:25	& I	& 300	& 0.70	\nl
97.04.13	& 00:32	& I	& 300	& 0.78	\nl
97.04.13	& 00:38	& I	& 300	& 0.75	\nl
97.04.13	& 00:46	& V	& 300	& 0.98	\nl
97.04.13	& 00:52	& V	& 300	& 1.00	\nl
97.04.13	& 00:59	& V	& 300	& 0.90	\nl
97.04.14	& 01:37	& V	& 600	& 0.96	\nl
97.04.14	& 01:49	& V	& 600	& 0.85	\nl
97.04.14	& 02:01	& V	& 600	& 0.83	\nl
\enddata
\end{deluxetable}

\clearpage
\newpage

\begin{deluxetable}{lcccc}
\tablenum{2}
\tablewidth{0pt}
\tablecaption{Comparative features and global properties of Antlia}
\tablehead{
\colhead{~} & \colhead{LGS 3} & \colhead{Pegasus} & \colhead{Antlia (center)} & \colhead{Antlia (outer)}}
\startdata
$N_{\rm red}$ & 104 & 1144 & 299 & 181 \nl
$N_{\rm blue}$ & 5 & 80 & 49 & 8 \nl
$N_{\rm blue}/N_{\rm red}$ & 0.05 & 0.07 & 0.16 & 0.04 \nl
$\bar\psi$ \hfill (M$_\odot$yr$^{-1}$) & $4\times 10^{-5}$ & $6.7\times 10^{-4}$ & [$1-2\times 10^{-4}$] & [$0.5-1\times 10^{-4}$] \nl
$\bar\psi_{\rm 1Gyr}$ \hfill (M$_\odot$yr$^{-1}$) & $1.5\times 10^{-5}$ & $6.2\times 10^{-4}$ & [$1.5-4\times 10^{-4}$] & [$<3\times 10^{-5}$] \nl
\enddata
\end{deluxetable}

\begin{deluxetable}{lcccc}
\tablenum{3}
\tablewidth{0pt}
\tablecaption{Comparative features and global properties of Antlia}
\tablehead{
\colhead{~} & \colhead{Antlia (center)} & \colhead{Antlia (outer)} & \colhead{LGS 3} & \colhead{Pegasus}}
\startdata
$\bar\psi/A$ \hfill (M$_\odot$yr$^{-1}$pc$^{-2}$) & $2-4\times 10^{-10}$ & $3-5\times 10^{-11}$ & $1.4\times 10^{-10}$ & $1.5\times 10^{-9}$ \nl
$\bar\psi_{\rm 1Gyr}/A$ \hfill (M$_\odot$yr$^{-1}$pc$^{-2}$) & $3-9\times 10^{-10}$ & $<10^{-11}$ & $5.5\times 10^{-11}$ & $1.4\times 10^{-9}$ \nl
$Z$ & \multispan{2}\hfill 0.0012 \hfill & 0.0014 & 0.002 \nl
Distance \hfill (Mpc) & \multispan{2}\hfill 1.32 \hfill & 0.77 & 0.95 \nl
Distance to LG \hfill (Mpc) & \multispan{2}\hfill 1.72 \hfill & 0.37 & 0.64 \nl
$M_{\rm V,0}$ & \multispan{2}\hfill $-10.3$ \hfill & $-9.28$ & $-12.55$ \nl
$L_{\rm V}$ \hfill (L$_{\sun}$) & \multispan{2}\hfill $1.2\times 10^6$ \hfill & $4.5\times 10^5$ & $9.3\times 10^6$ \nl
$M_{\rm \star}$ \hfill (M$_{\sun}$) & \multispan{2}\hfill $2-4\times 10^6$ \hfill & $4.8\times 10^5$ & $10^7$ \nl
$M_{\rm gas}$ \hfill (M$_{\sun}$) & \multispan{2}\hfill $(1.1\pm0.2)\times 10^6$ \hfill & $3.2\times 10^5$ & $6.5\times 10^6$ \nl
$M_{\rm tot}$ \hfill (M$_{\sun}$) & \multispan{2}\hfill $(4.1\pm0.8)\times 10^7$ \hfill & $1.8\times 10^7$ & $1.8\times 10^8$ \nl
\enddata
\end{deluxetable}

\clearpage
\newpage

\begin{figure}
\vspace{1cm}
\caption
{$I$ image of Antlia. The total field is
$7.8\arcmin\times 6.9\arcmin$ and the integration time 2400 s. The
central (A) and outer (B) regions in which the galaxy has been divided
for its study are shown. The two regions marked (C) have been used for
foreground subtraction. North is up, east is left. {\bf Note}: this figure is sent separately in gif format to reduce size}
\end{figure}

\clearpage
\newpage

\begin{figure}
\vspace{12cm}
\caption
{Finding chart of the resolved stars. Large filled
triangles represent Antlia stars bluer than the diagonal lines
in Fig. 4. Open circles are Antlia stars
redward of these lines. Small dots represent foreground stars. The
plain small dots inside regions A and B represent the stars that have been
eliminated from the CM diagrams of these regions in the field
subtraction process.  The axes are given in pixels ($0.259^{\arcsec}/{\rm
pixel}$).}
\includegraphics{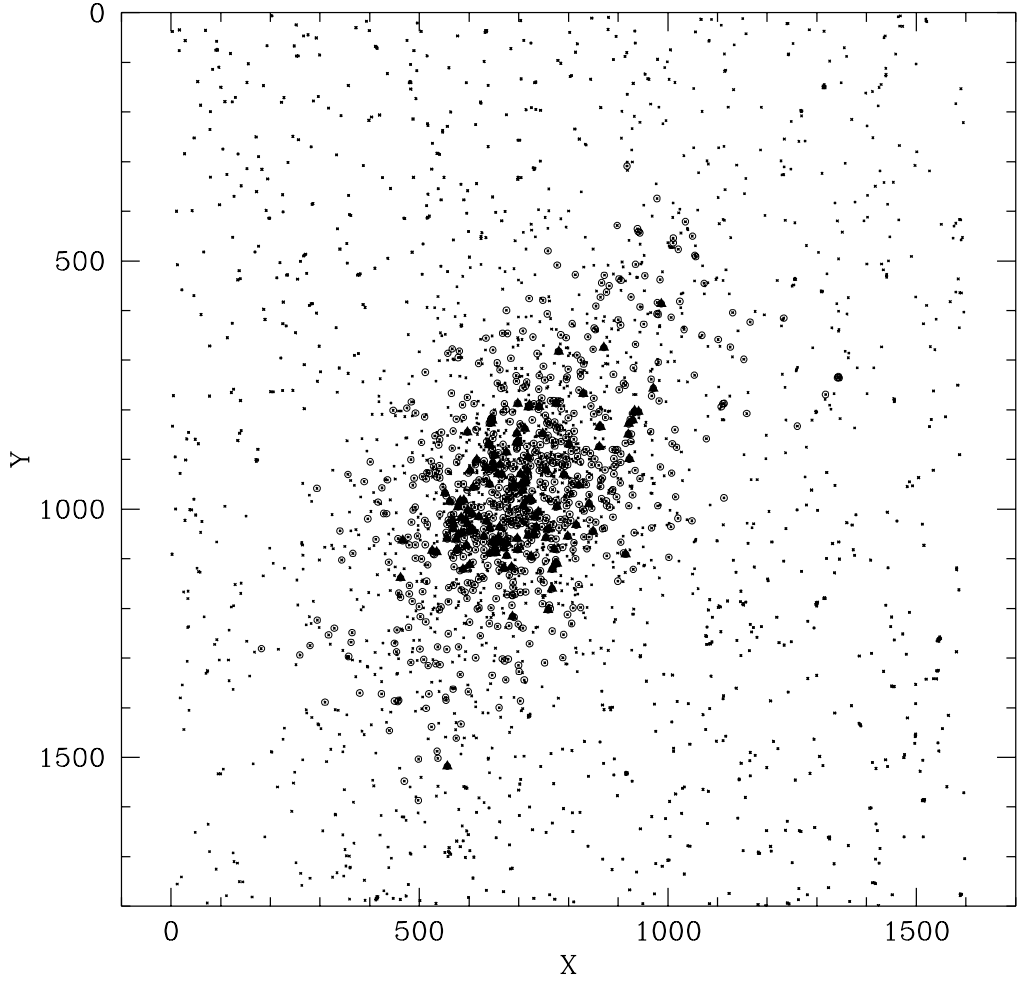}
\end{figure}

\clearpage
\newpage

\begin{figure}
\vspace{1cm}
\caption
{$I$ image of the central part of Antlia. The
total field is $2.2\arcmin\times 2.2\arcmin$ and the integration time
2400 s. The circle shows a possible HII region. North is up, east is
left. {\bf Note}: this figure is sent separately in gif format to reduce size}
\end{figure}

\clearpage
\newpage

\begin{figure}
\vspace{12cm}
\caption
{$I$ vs. $(V-I)$ CM diagrams for different regions
of the observed field. [a] Antlia central region (A of Fig. 1); [b]
Antlia outer region (B of Fig. 2); [c] and [d] external foreground
regions (C of Fig. 1) normalized to the areas covered in panels [a]
and [b] respectively; [e] and [f], the same diagrams shown in panels
[a] and [b] after subtraction of foreground stars. Diagonal lines show
the line used to discriminate blue and red stars in Fig. 2.}
\includegraphics{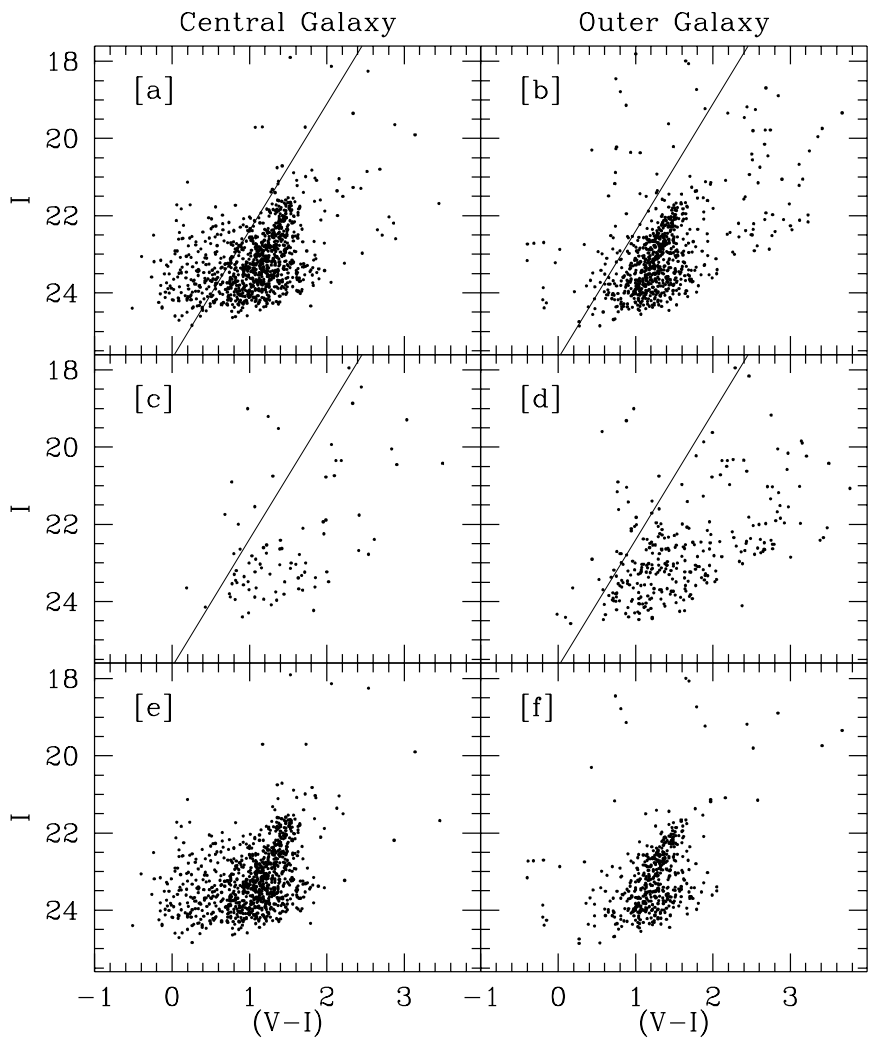}
\end{figure}

\begin{figure}
\vspace{12cm}
\caption
{$V$ surface brightness distribution as a
function of semimajor axis. It can be fitted by two exponential laws
with the indicated parameters.  The 1$\sigma$ error bars are calculated as the
quadrature sum of the uncertainties in the mean number of counts
in a given annulus and the uncertainty in the sky subtraction.}
\includegraphics{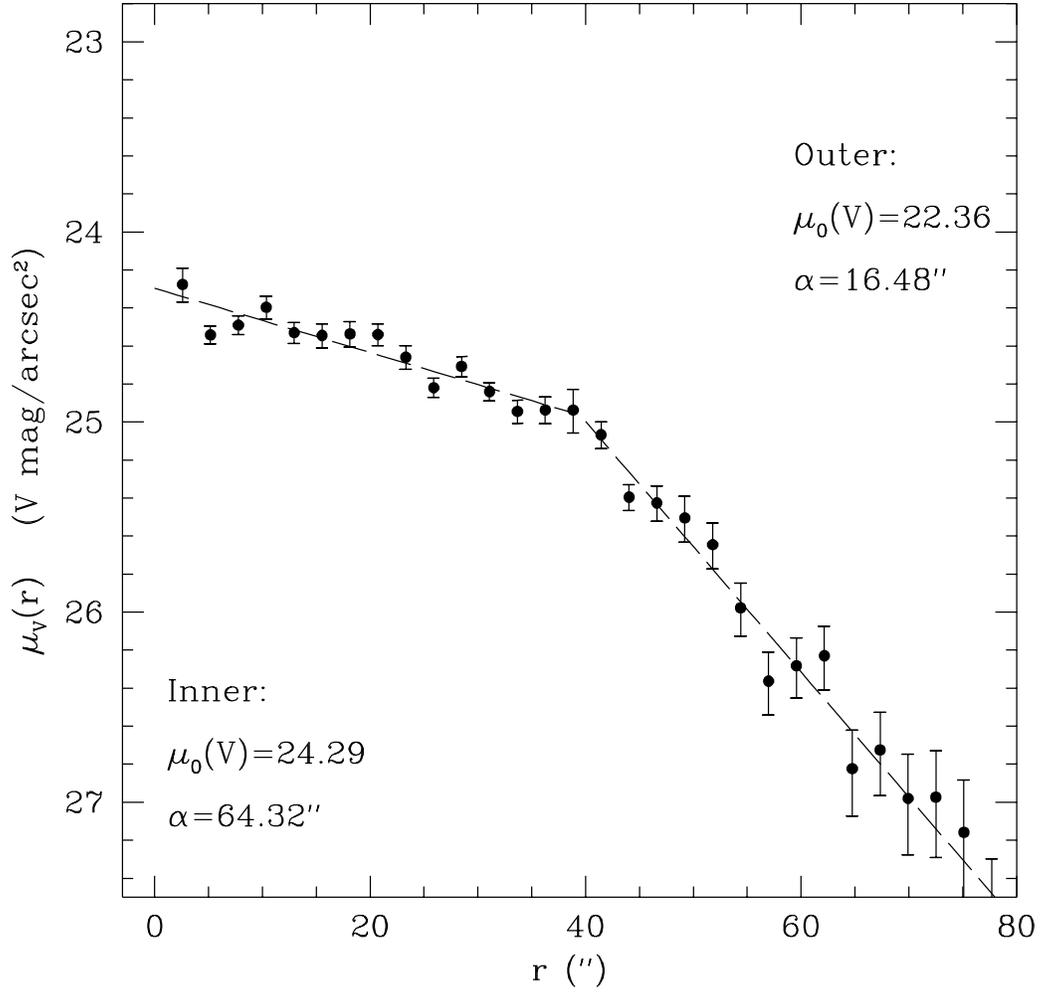}
\end{figure}

\begin{figure}
\vspace{12cm}
\caption
{$M_{\rm I,0}$ vs. $(V-I)_0$ CM diagrams of the
central and outer regions of Antlia compared to those of Pegasus and
LGS 3. An isochrone from the library of Padua (see Bertelli et
al. 1994) of 0.1 Gyr and $Z=0.001$ for Antlia and LGS and the same age
and $Z=0.004$ for Pegasus is overplotted.}
\includegraphics{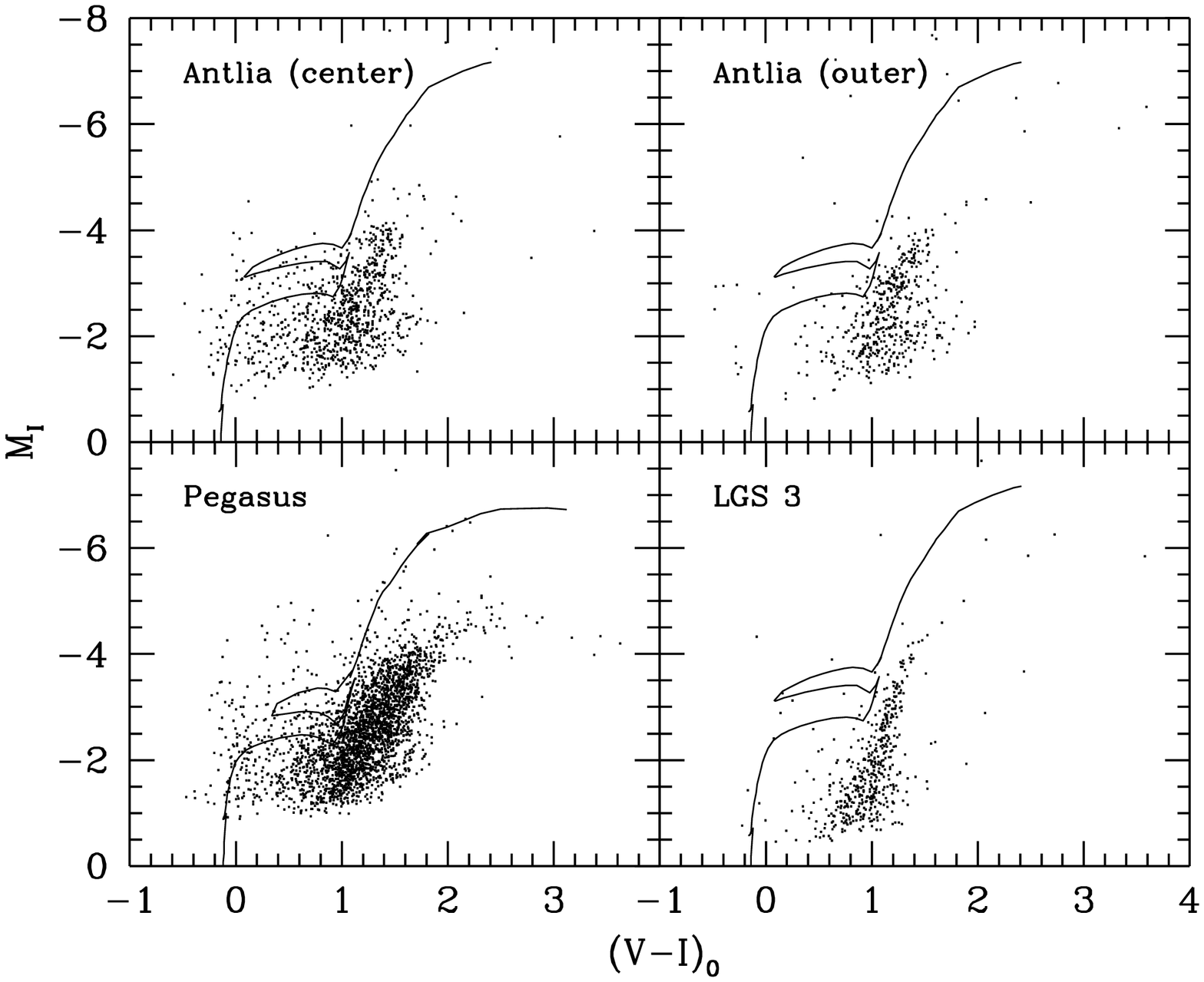}
\end{figure}


\begin{references}
\reference{lg} Aparicio, A., \& Gallart, C. 1994, in The Local Group: 
comparative and global properties, ESO Conference and Workshop Proceedings No. 
51, ed. A. Layden, R. C. Smith, \& J. Storm (Garching: ESO), 115
\reference{pegfot} Aparicio, A., \& Gallart, C. 1995, \aj, 110, 2105
\reference{peg} Aparicio, A., Gallart, C., \& Bertelli, G. 1997a, \aj, in press
\reference{lgs} Aparicio, A., Gallart, C., \& Bertelli, G. 1997b, \aj, in press
\reference{arp} Arp, H.C. \& Madore, B.F. 1987,  "A catalogue of southern peculiar galaxies
and associations. Vol. 1: Positions and descriptions" pag. 113, Cambridge University Press
\reference{azzopardi} Azzopardi, M. 1994, in The Local Group: 
comparative and global properties, ESO Conference and Workshop Proceedings No. 
51, ed. A. Layden, R. C. Smith, \& J. Storm (Garching: ESO), 129
\reference{arp} Arp, H.C. \& Madore, B.F. 1987,  "A catalogue of southern peculiar galaxies
and associations. Vol. 1: Positions and descriptions" pag. 113, Cambridge University Press
\reference{bertelli} Bertelli, G., Bressan, A., Chiosi, C., Fagotto, F., \& Nasi, 
E. 1994, \aaps, 106, 271
\reference{burstein} Burstein, D., \& Heiles, C. 1982, \aj, 87, 1165
\reference{caldwell} Caldwell, N., Armandroff, T. E., Seitzer, P., \& Da Costa, G. S. 1992, \aj, 103, 840
\reference{capa} Capaccioli, M., Piotto, G., \& Bresolin, F. 1992, \aj, 105, 1779
\reference{cardelli} Cardelli, J. A., Clayton, G. C., \& Mathis, J.S. 1989, \apj, 345, 245
\reference{coleman} Coleman, C. D., Wu, C.-C., \& Weedman, D. W. 1980, \apjs 43, 393
\reference{corwin} Corwin, H. G. jr., de Vaucoulers, A., \& de Vaucoulers, G. 1985, Southern Galaxy Catalog, University of Texas Monographs \#4
\reference{davidge92} Davidge, T. J. 1992, \apj, 397, 457
\reference{davidge} Davidge, T. J. 1993, \aj, 105, 1392
\reference{dacosta} Da Costa, G. S., \& Armandroff, T. E. 1990, \aj, 100, 162
\reference{dvau} de Vaucouleurs, G., de Vaucouleurs, A., Corwin, H.G., Buta, 
R.J., Paturel, G., \& Fouqu\'e, P. 1991, Third reference Catalog of Bright 
Galaxies (New York: Springer-Verlag)
\reference{durrant} Durrant, C. J. 1981, Landolt-B\"ornstein New Series, ed. 
by K.-H. Hellwege, Group VI, ed. K. Schaifers, \& H. H. Voigt,  vol.~2a,
(Berlin-Heidelberg: Springer), 82
\reference{feitz} Feitzinger, J. V., \& Galinski, Th. 1985, \aaps, 61, 503
\reference{fouque} Fouqu\'e, P, Bottinelli, L., Durrand, N., Gouguenheim, L, \& Paturel, G. 1990, \aaps, 86, 473
\reference{fukugita} Fukugita, M., Shimasaku, K., \& Ichikawa, T. 1995, \pasp 107, 945
\reference{n6822a} Gallart, C., Aparicio, A., V\'\i lchez, J. M.  1996a, \aj, 112, 1928
\reference{n6822b} Gallart, C., Aparicio, A., Bertelli, G., Chiosi, C. 1996b, \aj, 112, 1950
\reference{hartmann1} Hartmann, D. 1994, "The Leiden/Dwingeloo Survey of Galactic Neutral Hydrogen", Ph.D. Thesis, University of Leiden
\reference{hartmann2} Hartmann, D. 1997, "Atlas of Galactic Neutral Hydrogen", Cambridge University Press
\reference{hoffman} Hoffman, G. L., Salpeter, E. E., Farhat, B., Roos, T., Williams, H., \& Helou, G. 1996, \apjs, 105, 269
\reference{jobin} Jobin, M. \& Carignan, C. 1990, \aj, 100, 648
\reference{kar} Karachentseva, V. E., Prugniel, Ph., Vennink, J., Richter, G., 
Thuan, T., \& Martin, J. 1997, \aap, in press
\reference{landolt} Landolt, A. U. 1992, \aj, 104, 340
\reference{leoi} Lee, M. G., Freedman, W. L., Mateo, M., Thompson, I., \& Ruiz, M. T. 1993a, \aj, 106, 1420
\reference{trgb} Lee, M. G., Freedman, W. L., Madore, B. F. 1993b, \apj, 417, 553
\reference{lee} Lee, M. G. 1993, \apj, 408, 409
\reference{n185} Mart\'\i nez-Delgado, D., Aparicio, A., \& Gallart, C. 1997, in preparation
\reference{phoenix} Mart\'\i nez-Delgado, D., Aparicio, A., \& Gallart, C. 1998, in preparation
\reference{minniti} Minniti, D., \& Zijlstra, A. A. 1996, \apj, 463, L13
\reference{smecker} Smecker-Hane, T. A., Stetson, P. B., \& Hesser, J. E. 1994, \aj, 108, 507
\reference{allframe} Stetson, P. B. 1994, \pasp, 106, 250
\reference{fornax} Stetson, P. B. 1997, Baltic Astronomy, 6, 3
\reference{vanden} van den Bergh, S. 1994, \aj, 107, 1328
\reference{whitin} Whiting, A., Irwin, M \&, Hau, G. 1997, \aj (in press)
\end{references}
\end{document}